\documentstyle[epsf]{mn}

\newif\ifAMStwofonts

\newcommand{\kms}{km~s$^{-1}$}
\newcommand{\Msol}{M$_\odot$}
\newcommand{\Rsol}{R$_\odot$}
\newcommand{\Lsol}{L$_\odot$}
\newcommand{\g}{\log~g}
\newcommand{\te}{T_{\rm eff}}
\newcommand{\ld}{\lambda}
\newcommand{\lte}{\log~T_{\rm eff}}

\ifoldfss
  \ifCUPmtlplainloaded \else
    \NewTextAlphabet{textbfit} {cmbxti10} {}
    \NewTextAlphabet{textbfss} {cmssbx10} {}
    \NewMathAlphabet{mathbfit} {cmbxti10} {} 
    \NewMathAlphabet{mathbfss} {cmssbx10} {} 
  \fi
  \ifAMStwofonts
    \ifCUPmtlplainloaded \else
      \NewSymbolFont{upmath} {eurm10}
      \NewSymbolFont{AMSa} {msam10}
      \NewMathSymbol{\upi}     {0}{upmath}{19}
      \NewMathSymbol{\umu}     {0}{upmath}{16}
      \NewMathSymbol{\upartial}{0}{upmath}{40}
      \NewMathSymbol{\leqslant}{3}{AMSa}{36}
      \NewMathSymbol{\geqslant}{3}{AMSa}{3E}

    \fi
  \fi
\fi 

\ifnfssone
  \newmathalphabet{\mathit}
  \addtoversion{normal}{\mathit}{cmr}{m}{it}
  \addtoversion{bold}{\mathit}{cmr}{bx}{it}
  \newmathalphabet{\mathbfit} 
  \addtoversion{normal}{\mathbfit}{cmr}{bx}{it}
  \addtoversion{bold}{\mathbfit}{cmr}{bx}{it}
  \newmathalphabet{\mathbfss} 
  \addtoversion{normal}{\mathbfss}{cmss}{bx}{n}
  \addtoversion{bold}{\mathbfss}{cmss}{bx}{n}
  \ifAMStwofonts
    \ifCUPmtlplainloaded \else
      \UseAMStwoboldmath
      \makeatletter
      \new@mathgroup\upmath@group
      \define@mathgroup\mv@normal\upmath@group{eur}{m}{n}
      \define@mathgroup\mv@bold\upmath@group{eur}{b}{n}
      \edef\UPM{\hexnumber\upmath@group}
      \new@mathgroup\amsa@group
      \define@mathgroup\mv@normal\amsa@group{msa}{m}{n}
      \define@mathgroup\mv@bold\amsa@group{msa}{m}{n}
      \edef\AMSa{\hexnumber\amsa@group}
      \makeatother
      \mathchardef\upi="0\UPM19
      \mathchardef\umu="0\UPM16
      \mathchardef\upartial="0\UPM40
      \mathchardef\leqslant="3\AMSa36
      \mathchardef\geqslant="3\AMSa3E
    \fi
  \fi
\fi 

\ifnfsstwo
  \DeclareMathAlphabet{\mathbfit}{OT1}{cmr}{bx}{it}
  \SetMathAlphabet\mathbfit{bold}{OT1}{cmr}{bx}{it}
  \DeclareMathAlphabet{\mathbfss}{OT1}{cmss}{bx}{n}
  \SetMathAlphabet\mathbfss{bold}{OT1}{cmss}{bx}{n}
  \ifAMStwofonts
    \ifCUPmtlplainloaded \else
      \DeclareSymbolFont{UPM}{U}{eur}{m}{n}
      \SetSymbolFont{UPM}{bold}{U}{eur}{b}{n}
      \DeclareSymbolFont{AMSa}{U}{msa}{m}{n}
      \DeclareMathSymbol{\upi}{0}{UPM}{"19}
      \DeclareMathSymbol{\umu}{0}{UPM}{"16}
      \DeclareMathSymbol{\upartial}{0}{UPM}{"40}
      \DeclareMathSymbol{\leqslant}{3}{AMSa}{"36}
      \DeclareMathSymbol{\geqslant}{3}{AMSa}{"3E}
    \fi
  \fi
\fi 

\ifCUPmtlplainloaded \else
  \ifAMStwofonts \else 
    \def\upi{\pi}
    \def\umu{\mu}
    \def\upartial{\partial}
  \fi
\fi

\title{CD Tau: a detached eclipsing binary with a solar-mass companion}
\author[I. Ribas et al.]
       {Ignasi Ribas,$^1$ Carme Jordi$^1$ and Jordi Torra$^{1,2}$ \\
        $^1$Departament d'Astronomia i Meteorologia, Universitat de
           Barcelona, Av. Diagonal 647, E-08028, Barcelona, Spain\\
        $^2$Institut d'Estudis Espacials de Catalunya, Edif. Nexus-104, Gran 
           Capit\`a, 2-4, E-08034, Barcelona, Spain}
\date{Accepted ;
      Received ;
      in original form}

\pagerange{\pageref{firstpage}--\pageref{lastpage}}
\pubyear{1999}

\begin{document}

\maketitle

\label{firstpage}

\begin{abstract}

We present a detailed analysis of the detached eclipsing binary CD~Tau.
A large variety of observational data, in form of IR photometry, CORAVEL
radial velocity observations and high-resolution spectra, are combined
with the published light curves to derive accurate 
absolute dimensions and effective temperature of the components, as well as
the metal abundance of the system. We obtain: $M_{\rm A}=1.442(16)$~\Msol,
$R_{\rm A}=1.798(17)$~\Rsol, ${\te}_{\rm A}=6200(50)$~K, $M_{\rm B}=
1.368(16)$~\Msol, $R_{\rm B}=1.584(20)$~\Rsol~and ${\te}_{\rm B}=6200(50)$~K.
The metal content of the system is determined to be $[Fe/H]=+0.08(15)$~dex.

In addition, the eclipsing binary has a K-type close visual companion at about
10-arcsec separation, which is shown to be physically linked, thus
sharing a common origin. The effective temperature of the visual
companion (${\te}_{\rm C}=5250(200)$~K) is determined from synthetic
spectrum fitting, and its luminosity ($\log L/\mbox{\Lsol}=-0.27(6)$), and
therefore its radius ($R=0.89(9)$~\Rsol), are obtained from comparison with
the apparent magnitude of the eclipsing pair.

The observed fundamental properties of the eclipsing components are compared
with the predictions of evolutionary models, and we obtain good agreement for
an age of 2.6~Gyr and a chemical composition of $Z=0.026$ and $Y=0.26$.
Furthermore, we test the evolutionary models for solar-mass stars and we
conclude that the physical properties of the visual companion are very
accurately described by the same isochrone that fits the more massive
components.

\end{abstract}


\section{Introduction}

Double-lined eclipsing binaries are the only objects, apart from the Sun,
for which fundamental and simultaneous determinations of masses and radii
can be obtained. These determinations are possible through the analysis of
spectroscopic data in the form of radial velocity curves, and from modelling
the photometric data in light curves. In addition, if we consider detached 
double-lined eclipsing binaries, no significant mass transfer has occurred 
between the components, since they are smaller than their respective Roche 
lobes. In such a case, the mutual interaction can be safely neglected and the 
components can be assumed to evolve like single, individual stars. Therefore, 
these objects yield simultaneous absolute dimensions for two single
stars that are supposed to have a common origin both in time and 
chemical composition. In this situation, evolutionary models should 
be able to predict the same age for both components for a certain chemical 
composition.

Eclipsing binaries that are members of physically bound multiple systems 
provide additional constraints for the analysis of evolutionary models. All 
the information that can be extracted for the companion/s (effective 
temperature, magnitude difference, mass, \dots) should be also fitted by the 
same isochrone that fits the eclipsing binary pair. Not many studies have made 
use of this possibility up to date, mainly because of the scarce amount of 
information available for the additional stellar companions.

CD Tau (HD~34335, HIP~24663) is a bright eclipsing binary ($V_{\rm 
max}=6.75$) composed of two similar F6~V stars, which was suggested by
Chambliss (1992) as a possible member of a triple system. It has a K-type 
close visual companion (CD~Tau~C) at 9.98~arcsec (ESA 1997) with a magnitude 
of $V=9.9$. The angular distance between the eclipsing pair (CD~Tau~AB) and 
CD~Tau~C is large enough to allow the possibility of obtaining relevant 
astrophysical information for all three components separately.

Evolutionary models for stars with $M\ga1.2$~\Msol~are known to provide a
reasonably good description of the observed stellar properties, at least
within the main sequence. The relatively large amount of eclipsing binary and 
open cluster data in this mass range (Andersen 1991; Pols et al. 1998) has 
been of remarkable help for reaching such an agreement. Nevertheless, recent 
studies (Popper 1997; Clausen, Helt \& Olsen 1999) have pointed out that 
current stellar evolutionary models are unable to appropriately describe the 
observed properties of low-mass eclipsing binaries (0.7--1.1~\Msol). The 
authors have found that the models for the less massive stars seem to predict 
radii that are too small ($\g$ too large) and effective temperatures that are 
too high when compared to observational data. Clausen et al. (1999) suggest 
several possible causes of the discrepancy: observational problems, lack of 
chemical composition determinations and the inaccuracies of the physics of 
the models, in particular the mixing length parameter. In any case, the number 
of systems in this mass range is still very small to reach statistically 
significant conclusions. CD~Tau~C, which is apparently a sub-solar mass star, 
may yield useful information for evolutionary model testing and could also 
help in the solution of the dilemma indicated by Popper (1997) and Clausen 
et al. (1999).

In this paper, we give a detailed description of the available observational
data on CD~Tau and their analysis for the determination of the fundamental 
properties of the three components. Moreover, the existence of a physical link 
between CD~Tau~AB and CD~Tau~C is proved. We finally compare the resulting 
parameters with the evolutionary model predictions in order to assess their 
ability to reproduce the observations. 

\section{Modelling the radial velocity and light curves} \label{lcrv}

A radial velocity curve of CD~Tau was published by Popper (1971), but the
quality of the data and the poor phase coverage do not allow an accurate
determination of the masses. Therefore, new radial velocity observations were
secured with the CORAVEL scanner (Baranne, Mayor \& Poncet 1979) attached to 
the 1-m telescope of the Observatoire de Gen\`eve and located in the 
Observatoire d'Haute Provence (France). Two different observational campaigns 
(1997 February and October) allowed us to collect 48 and 46 measurements for 
the primary and secondary components, respectively, and 4 measurements for the 
visual companion. Heliocentric corrections to both Julian date and radial 
velocity were applied, and the correlation dips were analysed by considering 
colour dependence (see Baranne et al. 1979; Duquennoy, Mayor \& Halbwachs 
1991). The final radial velocity measurements for all components, together 
with their errors, are listed in Table \ref{tab:rvCD}.

\begin{table}
  \begin{center}
    \caption{CORAVEL radial velocity observations of CD~Tau~A, B and C.}
    \label{tab:rvCD}
{\footnotesize
    \begin{tabular}{rrcrrc}
\hline
\multicolumn{1}{c}{HJD --}&\multicolumn{1}{c}{$v_{\rm r}$}&\multicolumn{1}{c}{$\sigma_{v_{\rm r}}$}&
\multicolumn{1}{c}{HJD --}&\multicolumn{1}{c}{$v_{\rm r}$}&\multicolumn{1}{c}{$\sigma_{v_{\rm r}}$}\\
2450000  & \multicolumn{2}{c}{\scriptsize (\kms)} &
2450000  & \multicolumn{2}{c}{\scriptsize (\kms)} \\
\hline
\multicolumn{6}{c}{\normalsize A component} \\
\hline
498.285 &    65.0&1.4&500.414&$-$106.1&1.1\\
498.305 &    65.6&1.3&501.281&    23.7&1.2\\
498.316 &    68.8&1.2&501.289&    26.1&1.0\\
498.332 &    67.6&1.4&501.309&    28.3&1.1\\
498.344 &    67.7&1.2&501.316&    29.8&1.2\\
498.355 &    69.8&1.1&501.340&    32.0&1.1\\
498.371 &    68.2&1.0&501.352&    34.4&1.2\\
498.387 &    67.6&1.1&501.441&    45.3&1.2\\
498.398 &    68.0&1.1&502.270&    35.5&1.0\\
498.414 &    70.3&1.3&502.277&    34.6&1.1\\
498.434 &    65.3&1.3&502.305&    31.7&1.1\\
499.367 & $-$50.7&1.7&505.262&    67.1&1.3\\
499.379 & $-$51.8&1.3&505.270&    66.2&1.5\\
499.395 & $-$54.7&1.2&505.289&    67.1&1.0\\
499.414 & $-$60.3&2.4&505.305&    65.8&1.1\\
499.434 & $-$62.4&1.2&505.363&    66.0&1.0\\
500.273 &$-$123.9&1.1&729.576& $-$58.8&1.4\\
500.281 &$-$122.1&1.3&729.588& $-$61.3&1.2\\
500.293 &$-$122.3&1.3&729.595& $-$62.4&1.3\\
500.320 &$-$119.5&1.0&730.587&$-$104.3&1.2\\
500.332 &$-$122.6&1.3&732.678&  $-$0.2&1.1\\
500.336 &$-$116.4&1.3&733.631&$-$123.7&1.0\\
500.391 &$-$114.8&1.3&736.626& $-$86.5&1.2\\
500.402 &$-$112.5&1.1&737.627& $-$85.8&1.1\\
\hline
\multicolumn{6}{c}{\normalsize B component} \\
\hline
498.293&$-$128.7&1.4&500.422&    56.0&1.4\\
498.312&$-$131.8&1.3&501.273& $-$86.2&1.5\\
498.324&$-$134.1&1.6&501.285& $-$88.8&1.6\\
498.340&$-$131.9&1.5&501.312& $-$92.3&1.6\\
498.352&$-$131.4&1.3&501.320& $-$92.7&1.5\\
498.359&$-$130.4&1.4&501.344& $-$97.0&1.5\\
498.379&$-$131.9&1.9&501.355& $-$96.2&1.6\\
498.391&$-$131.2&1.3&501.449&$-$108.2&1.5\\
498.406&$-$127.9&1.3&502.266& $-$97.8&1.5\\
498.426&$-$129.8&1.4&502.273& $-$97.6&1.5\\
498.441&$-$131.0&1.4&502.281& $-$95.9&1.6\\
499.371& $-$10.1&2.0&502.301& $-$94.8&1.5\\
499.387&  $-$4.8&1.7&505.266&$-$132.7&1.6\\
499.402&     2.0&1.5&505.270&$-$131.2&1.7\\
499.426&     5.4&1.7&505.281&$-$133.0&1.4\\
499.441&     5.6&1.6&505.297&$-$132.0&1.6\\
500.277&    65.1&1.3&729.581&     7.6&1.5\\
500.285&    67.1&1.6&729.602&     9.8&1.5\\
500.297&    64.6&2.1&730.601&    53.2&1.7\\
500.328&    62.2&1.7&732.682& $-$49.8&1.9\\
500.336&    63.2&1.6&733.635&    71.3&1.4\\
500.340&    63.5&1.4&736.620&    35.4&1.7\\
500.398&    59.8&1.4&737.627&    32.9&1.5\\
\hline
\multicolumn{6}{c}{\normalsize C component} \\
\hline
498.297&$-$27.6 & 1.5 &500.346& $-$26.9 & 1.6\\
500.289&$-$29.1 & 1.8 &501.289& $-$27.7 & 1.4\\   
\multicolumn{6}{c}{Mean value: $-27.8\pm0.8$}    \\
\hline
    \end{tabular}}
  \end{center}
\end{table}

\begin{figure}
\leavevmode
\epsfxsize=\hsize
\epsfbox{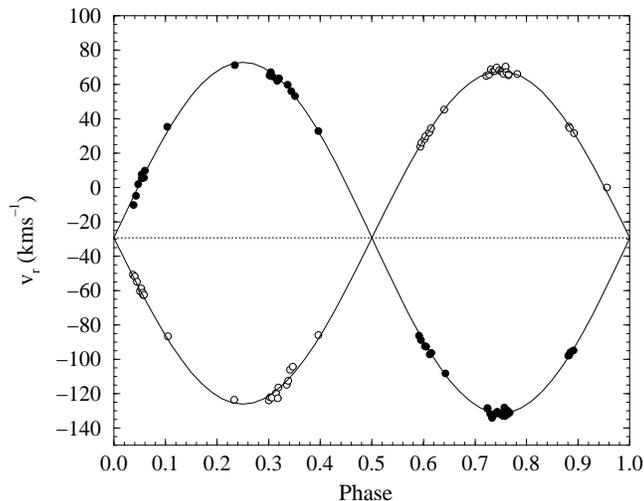}
\caption{Radial velocity curve fit to CD~Tau~AB CORAVEL observations
from 1997 February and October. Circular orbit is adopted. The r.m.s. 
residual of the fit is 0.3~\kms.}
\label{fig:rvCD}
\end{figure}

The solution of a radial velocity curve for a circular orbit, such as that of 
CD~Tau~AB, is a conceptually very simple problem. The only free parameters 
are the two velocity semi-amplitudes ($K_{\rm A}$ and $K_{\rm B}$), the 
systemic velocity ($\gamma$) and a phase offset. The ephemeris for CD~Tau,
for both radial velocity and light curve solutions, was adopted as 
$\mbox{Min I} = 2441619.4075 + 3\fd435137 \, E$ (Kholopov et al. 1987). We used 
the SBOP program
(created by Dr. P.B. Etzel in 1978 and later revised several times) and 
adopted the Lehmann-Filh\'es method (Lehmann-Filh\'es 1894; Underhill 1966) for 
simultaneous solution of a double-lined radial velocity curve. The parameters 
resulting from this analysis are shown in Table \ref{tab:CDTau} and a plot of 
the observed radial velocities and the best-fitting curve is presented in 
Fig. \ref{fig:rvCD}. The r.m.s. residual of the fit is as small as 0.3~\kms. 
The errors of the parameters were conservatively adopted as twice the 
standard errors provided by the SBOP program. Our results show good agreement 
with the analysis of the previously published radial velocity curve of 
Popper (1971), although the curve coverage and the individual accuracy of the 
measurements are significantly better in our study, leading to smaller 
formal errors.
 
Three complete photoelectric light curves of CD~Tau have been observed
to our knowledge by Srivastava (1976), Wood (1976) and G\"ulmen et al. (1980).
The best data are those of Wood, obtained in the Str\"omgren system, but,
unfortunately, the individual measurements were not published and are not
available. Thus, for this work, we made use of the observations of G\"ulmen
et al. with 292 measurements in each of $B$ and $V$ filters.

\begin{figure}
\leavevmode
\epsfxsize=\hsize
\epsfbox{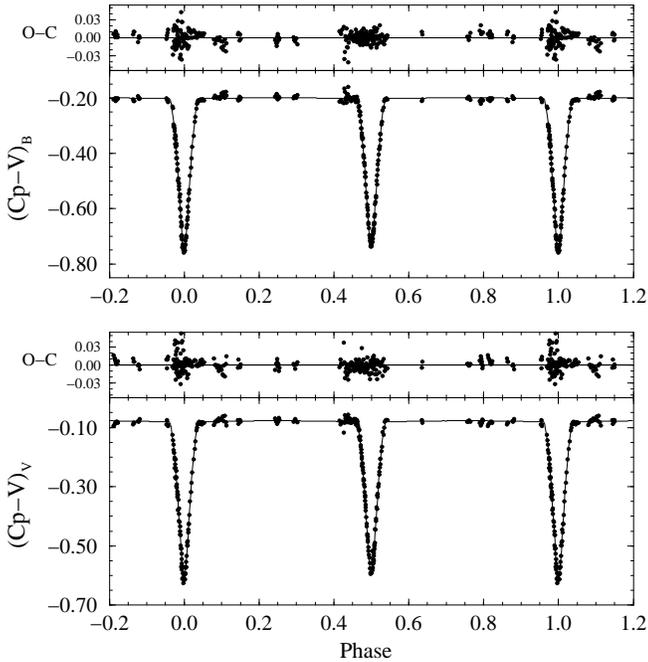}
\caption{Light curve fit to the observed $B$ and $V$ Comparison$-$Variable
(Cp$-$V) differential photometry of CD~Tau. Also shown are the 
Observed$-$Computed (O$-$C) residuals. Solutions obtained with WD program 
applied to the observations of G\"ulmen et al. (1980).}
\label{fig:lcCD}
\end{figure}

The analysis of light curves was done by using an improved version of the 
Wilson \& Devinney (1971, hereafter WD) program (Milone, Stagg \& Kurucz 1992; 
Milone et al. 1994) that 
includes Kurucz (1991, 1994) ATLAS9 atmosphere models. As suggested by the 
uncomplicated shape of the light curve (Fig. \ref{fig:lcCD}), we chose a 
detached configuration with coupling between luminosity and temperature when 
running the solutions. Reflection and proximity effects were considered for 
the sake of completeness, but their effect is expected to be negligible for 
such a well-detached system. The bolometric albedos were set to 0.5 (typical 
for convective atmospheres) and the gravity darkening exponents were adopted 
as 0.28 and 0.30 (A and B component, repectively) as deduced from the new 
computations of Claret (1998). The mass ratio $q$ was fixed to the 
spectroscopic value of 0.948, circular orbit was adopted and the temperature 
of the primary star was set to 6200~K (see Sect. \ref{tef}). Finally, both 
components were assumed to rotate synchronously with the orbital period.

Both $BV$ light curves were solved for simultaneously in order to
achieve a single, mutually consistent solution. The free parameters in
the light curve fitting were: the phase offset, $\phi_{\circ}$, the 
inclination, $i$, the temperature of the secondary, $T_{\rm B}$, the 
gravitational potentials, $\Omega_{\rm A}$ and $\Omega_{\rm B}$, the 
monochromatic luminosity of the primary, $L_{\rm A}$, and a third light 
contribution, $F_3$. 

An automatic procedure was employed to run the WD program. At each iteration, 
the differential corrections were applied to the input parameters to build the 
new set of parameters for the next iteration. The result was carefully 
checked in order to avoid possible unphysical situations (like e.g. Roche 
lobe filling in detached configuration). A solution was defined as the set 
of parameters for which the differential corrections suggested by the WD 
program were smaller than the standard errors during three consecutive 
iterations.  However, when a solution was found, the program did not stop. 
Instead, it was kept running in order to test the stability of the solutions, 
to evaluate their scatter and to check for possible spurious solutions. In 
the case of CD~Tau, several runs done from different initial conditions 
converged to a stable minimum. We finally adopted the parameters that are 
listed in Table \ref{tab:CDTau}. 

\begin{table*}
  \begin{center}
    \caption{Orbital and stellar parameters for CD~Tau~A and B derived 
from the analysis of the light and radial velocity curves, standard
photometry and spectroscopy. The numbers in parentheses are the adopted 
errors affecting the last digits of each parameter.}
    \label{tab:CDTau}
    \begin{tabular}{lll}
      \hline 
             \multicolumn{3}{c}{Radial velocity curve solution}\\
      \hline
$K_{\rm A} = 96.8(5)$ \kms & $q=\frac{M_{\rm B}}{M_{\rm A}}=0.948(7)$ & 
$\gamma=-29.3(3)$ \kms \\
$K_{\rm B}=102.1(5)$ \kms & $a=13.52(7)$ \Rsol &\\
      \hline 
             \multicolumn{3}{c}{Light curve solution}\\
      \hline 
$P=3.435137^{\rm a}$ days & $e=0^{\rm a}$ &
$ \left. \frac{L_{\rm B}}{L_{\rm A}} \right|_{B}$ = 0.770(22)\\
$r_{\rm A}$ = 0.1330(10) & $i=87\fdg7(3)$ & 
$ \left. \frac{L_{\rm B}}{L_{\rm A}} \right|_{V}$ = 0.772(23)\\
$r_{\rm B}$ = 0.1172(13) & $\frac{{\te}_{\rm B}}
{{\te}_{\rm A}}$ = 0.999(1) & $F_3|_B = 0.052(15)^{\rm b}$\\
&&$F_3|_V = 0.066(15)^{\rm b}$\\
      \hline
             \multicolumn{3}{c}{Photometry and spectroscopy}\\
      \hline
$\overline{T}_{\rm eff}=6200(50)$ K & $[Fe/H]=+0.08(15)$ dex
& $E(B-V)=0.0$ mag\\
${v_{\rm rot}}_{\rm A}=28(3)$~\kms & ${v_{\rm rot}}_{\rm B}=26(3)$~\kms &\\
      \hline
             \multicolumn{3}{c}{Stellar physical properties}\\
      \hline
\multicolumn{3}{c}{\(\begin{array}{lll}
M_{\rm A} = 1.442(16)~{\rm M}_{\odot}  &\mbox{\hspace{2cm}}& M_{\rm B} = 1.368(16)~{\rm M}_{\odot}  \\
R_{\rm A} = 1.798(17)~{\rm R}_{\odot} && R_{\rm B} = 1.584(20)~{\rm R}_{\odot} \\
\log g_{\rm A} = 4.087(10)~{\rm dex} && \log g_{\rm B} = 4.174(12)~{\rm dex} \\
{\te}_{\rm A} = 6200(50)~{\rm K} && {\te}_{\rm B} = 6200(50)~{\rm K}
\end{array}\)
}\\
      \hline
    \end{tabular}

\vspace{1mm}
$^{\rm a}$ Adopted, $^{\rm b}$ Phase dependent quantity: reference phase = 
$0\fp25$
  \end{center}
\end{table*}

A especially conflictive parameter is the third light since possible strong 
correlations with other free parameters may drive us to a wrong solution.
The correlation matrix was checked for this possibility and we noticed weak 
correlations between $F_3$ and all the remaining parameters but the 
inclination, for which the correlation coefficient was close to 0.9. This is a 
predictable result since the presence of third light affects the relative 
depths of the eclipses. Actually, the inclusion of $F_3$ as a free parameter 
in the solution of the light curve should only obey to an external evidence 
of some light excess. For this system, the situation is clear: the photometric 
observations of G\"ulmen et al. (1980) included the visual companion inside 
the diaphragm so that some amount of extra light is certainly present. Since 
we know that the $V$ magnitude difference is about 3~mag and that the spectral 
type is later than that of the eclipsing pair components ($\Delta (B-V) 
\approx0.4$, from Tycho photometry in the Hipparcos and Tycho catalogues, 
ESA 1997), we expect an excess of light of about 6.3 per cent in $V$ and 
4.4 per cent in 
$B$. These figures are actually very close to the values derived from the 
light curve analysis, and the colour behaviour is also reproduced, thus 
giving additional physical meaning to the best-fitting parameters obtained.

The uncertainties in the parameters presented in Table \ref{tab:CDTau} were
carefully evaluated by means of two different approaches. On the one hand, the
WD program yields the standard error associated to each adjusted parameter.
On the other hand, the error can also be estimated as the r.m.s. scatter of
{\em all} the parameter sets corresponding to the iterations between the first
solution and the last iteration (about 100). We shall mention that this is, 
in general, larger than the r.m.s. scatter of the solutions alone, i.e. those 
that fulfill the criterion previously described. We conservatively adopted 
the uncertainty as twice whichever of the two error estimates was the largest.
The r.m.s. scatters of the residuals in the light curve fit are 0.011~mag 
($n=292$) for both $B$ and $V$. Fig. \ref{fig:lcCD} shows the light curve 
fit to the observed $B$ and $V$ differential photometry and the corresponding 
residuals, where no systematic trends are observed.

We shall compare our analysis with that of Wood (1976), G\"ulmen et al. (1980)
and Russo et al. (1981). Regarding the first two works, significant differences 
exist in the inclination and in the fractional radii. The inclination is 
0.8--1$^{\circ}$ larger in our solution, $r_{\rm A}$ is also 0.010 larger 
in our case, but $r_{\rm B}$ is about 0.005 smaller, the latter implying a 
large change in the ratio of radii. However, our computed elements are in 
very good agreement with those of Russo et al. (1981). They also employed the 
WD program (although an earlier version) to analyse the light curves, while the 
other authors used Wood's model. Thus, the reason for the discrepancy in the 
solutions is probably related to intrinsic differences of the models. The 
radii of Wood (1976) and G\"ulmen et al. (1980) lead to a physically unlikely 
situation for a detached system, with a secondary component too evolved when 
compared to the primary ($\g_{\rm A} \ga \g_{\rm B}$). 

As a brief summary, our final solution shows a system composed of two similar 
stars of about 1.4~\Msol~in a moderately evolved stage ($\g\approx4.1$), 
although still in the main sequence. The remarkable point is that the absolute 
dimensions of the components are known to a precision of about 1 per cent. 

\section{Effective temperature determination} \label{tef}

\begin{table*}
  \begin{center}
    \caption{Str\"omgren (from Hauck \& Mermilliod 1998) and IR-band 
(this work) observations for CD~Tau~AB. The $JHK$ measurements were taken at 
quadrature, when full light contribution from both components is present.
$n$ is the number of averaged photometric measurements.}
    \label{tab:tefCD}
{\footnotesize
    \begin{tabular}{ccccccccc}
\hline
$V$ & $(b-y)$ & $m_1$ & $c_1$ & $\beta$ & $n$ & $\te$ &
{\scriptsize $E(b-y)$} & {\scriptsize  $\g$} \\
\hline
6.742$\pm$0.007& 0.322$\pm$0.003&0.157$\pm$0.003&
0.456$\pm$0.003&2.627&4&6200& 0.0 & 3.9 \\
\hline
\hline
$J$ & $H$ & $K$ & $n$ & ${\te}_J$ & ${\te}_H$ & ${\te}_K$ &
{\scriptsize  $[Fe/H]^{\rm a}$} & {\scriptsize  $\g^{\rm a}$} \\
\hline
5.791$\pm$0.007&5.579$\pm$0.008&5.530$\pm$0.006&6&6185&6230&6211&0.08&4.1 \\
\hline
    \end{tabular}

\vspace{1mm}
$^{\rm a}$ Adopted.}
  \end{center}
\end{table*}

Str\"omgren-band and IR-band photometry were used for the determination of
the ``mean'' effective temperature of CD~Tau~AB. Standard Str\"omgren indices 
for CD~Tau~AB were taken from the catalog of Hauck \& Mermilliod (1998) and 
are shown in Table \ref{tab:tefCD}. We adopted the extinction correction 
calibration of Crawford (1975) and the photometric grids of Napiwotzki (1998), 
based on Kurucz ATLAS9 atmosphere models. The application of these calibrations 
indicated a negligible amount of interstellar extinction and yielded a value 
of 6200~K for the effective temperature, as shown in Table \ref{tab:tefCD}. 

In order to better constrain the effective temperature determination, several 
measurements of CD~Tau in the IR bands were obtained in the 1~m Carlos 
S\'anchez Telescope (Tenerife, Spain). The observations were made on 1998 
February 19 and 24, and April 1 and 3, when the eclipsing component was near 
quadrature. The reduction of the data and the transformation to the $J$, $H$ 
and $K$ standard filters was done by using the procedure explained in Masana 
(1999). The Infra-Red Flux Method (IRFM, Blackwell et al. 1990) was employed 
to obtain an accurate determination of the effective temperature, by comparing
the observed flux distribution with Kurucz ATLAS9 model atmospheres. A $\g$ of 
4.1 and $[Fe/H]=+0.08$ (see Sect. \ref{feh}) were adopted as input parameters 
in the analysis, although their influence is almost negligible. The resulting 
values of the effective temperature obtained at each passband are listed in 
Table \ref{tab:tefCD}. 

As it can be seen, the mutual agreement between the three $JHK$ determinations 
is very good, and they are almost identical to the independent value obtained 
from Str\"omgren photometry. Therefore, we can confidently adopt a 
$\overline{T}_{\rm eff} = 6200\pm50$~K, where the error includes both the 
variance of the average for all filters and the calibration error of
the IRFM method (see Alonso, Arribas \& Mart\'{\i}nez-Roger 1996).

\section{Chemical composition} \label{feh}

Two high-resolution spectra (0.15 and 0.12~\AA/pix) in different 
spectral regions were obtained with the aim of determining the atmospheric 
chemical composition of CD~Tau~AB. The instrument used was an echelle 
spectrograph attached to the 74-inch (1.9~m) telescope of Mount Stromlo 
Observatory (Australia). The spectra were obtained on 1997 October 17 and 19, 
when the eclipsing component was at phase 0\fp87 and 0\fp44, respectively. 
One of the spectra, with an integration time of 900~s, spans 5730 to 
6045~\AA~(around the Na\,{\sc i}d line of the interstellar medium), whereas the other 
one covers the range from 3830 to 4070~\AA~(including the strong lines of 
Ca\,{\sc ii}~K and H) with an integration time of 600~s. The $S/N$ ratio is of the 
order of 60 and 20, respectively. As a preliminary reduction, the raw images 
were corrected for bias, dark current and flat field using the standard 
procedures (see Jordi et al. 1995) and then wavelength calibrated. Radial 
velocity corrections were applied to refer the wavelength scale to the 
heliocentric reference frame and, finally, the flux scale was normalized 
to unity.

The element abundances were determined by fitting of synthetic spectra to 
our observed spectra. Unfortunately, their $S/N$ ratios (especially for the 
one around Ca\,{\sc ii} lines) are too low to allow a very accurate determination of 
the chemical composition (see Cayrel de Strobel 1985; Edvardsson et al. 1993) 
but still enough for a preliminary estimation. 

For the synthetic spectrum fitting, we adopted Kurucz ATLAS9 models and the 
set of programs developed by Dr. I. Hubeny ({\sc synspec}, {\sc rotin} and 
{\sc synplot}). Nevertheless, we modified these programs by considering 
the double-lined nature of our spectra, in such a way that they use two 
atmosphere models, two rotational velocities, two turbulent velocities and 
two radial velocities as input data. The two resulting synthetic spectra are 
subsequently co-added. Since we are dealing with an eclipsing binary system, 
the surface gravities and the temperature ratio obtained by modelling 
the light and radial velocity curves provide us with useful constraints. 
The ``mean'' system $\g$ is 4.1~dex, so we adopted Kurucz atmosphere models at 
a $\g$ of 4.0~dex. Nevertheless, several tests with $\g=4.5$~dex showed a very 
weak influence of this parameter on the synthetic spectra. The effective 
temperatures are strongly correlated with the atmospheric chemical composition 
determination. Thus, it is necessary to adopt accurate and mutually 
consistent effective temperatures. From the very precise $\te$ determinations 
through IR and Str\"omgren photometry, we adopted atmosphere models computed 
at $\te=6200$~K for both components. Since the temperatures are identical, 
but not the luminosities (i.e. the radii), a factor of 0.77 (see Table 
\ref{tab:CDTau}) was applied to the flux output of the secondary component 
when co-adding the synthetic spectra.

The radial velocity shifts of each component were computed from the radial 
velocity curve at the actual phases and introduced as a fixed parameter 
(42~\kms~and $-$103~\kms~for the Na\,{\sc i}d-region spectrum and $-$65~\kms~and 
8~\kms~for the Ca\,{\sc ii}-region spectrum). The rotational velocities of each 
component were initially adopted as the synchronization values, 26~\kms~and 
23~\kms, and were subsequently changed to 28~\kms~and 26~\kms~in order to 
obtain the best fit to the line profiles. The error of these rotational 
velocities is estimated to be about 3~\kms.

Taking into account all the adopted input parameters, the spectrum was studied 
in small intervals 20~\AA~wide. From several fitting tests at different 
Fe-peak element abundances ($-$0.1~dex, $-$0.05~dex, 0.0~dex, +0.05~dex, 
+0.1~dex and +0.15~dex) we deduced that the observed spectrum is best 
reproduced with a slightly super-solar composition. We estimate a value 
of $[Fe/H]=+0.08\pm0.15$~dex. The lack of lines of other elements does not 
allow abundance determinations, but there exists evidence of a slight 
underabundance of Na and a slight overabundance of Si, however, with low 
reliability. The parameters derived from the spectrum analysis are listed in 
Table \ref{tab:CDTau}. 

In general, the spectrum covering 5730--6045~\AA~was found to be of little 
help to obtain a Fe abundance determination since there are no strong lines 
of Fe-peak elements, although we obtained a good fit to the stellar Na\,{\sc i}d 
line. Moreover, the reddest zone of the spectrum ($\ld > 5970$~\AA) was 
strongly contaminated by non-stellar features.

\begin{figure}
\leavevmode
\epsfxsize=\hsize
\epsfbox{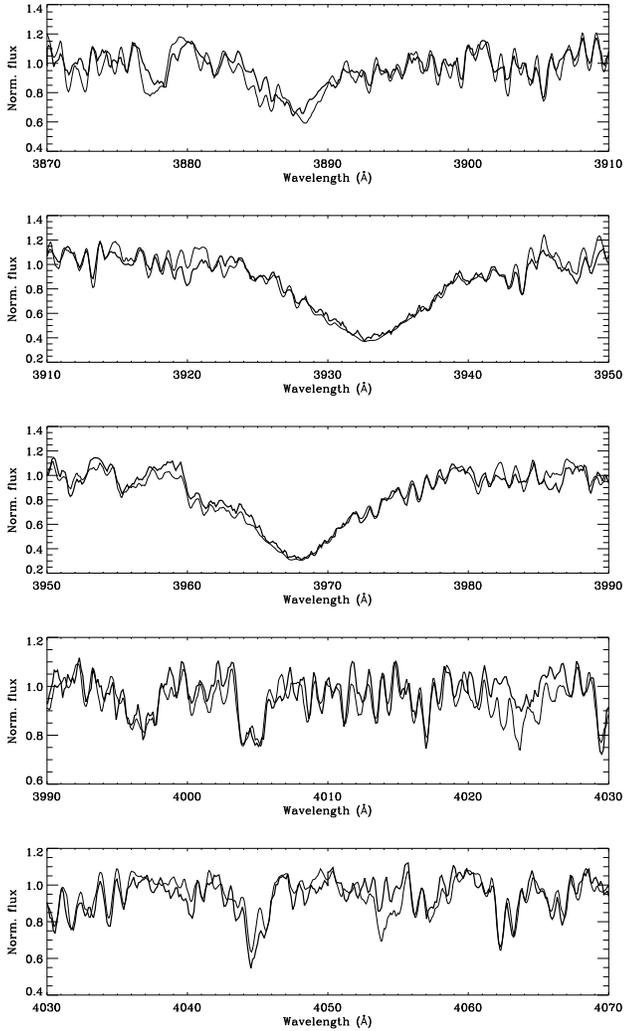}
\caption{Fit to an echelle spectrum of CD~Tau~AB (double-lined, phase = 0\fp44) 
from 3910 to 4070~\AA~obtained at Mt. Stromlo (Australia) on 1997 October 19. 
The thick line represents the observed spectrum whereas the thin line 
is the best-fitting synthetic spectrum computed using Kurucz ATLAS9 fluxes. 
The parameters used for the synthetic spectrum computation as well as some 
comments to individual features can be found in the text.}
\label{fig:spCD}
\end{figure}

The spectrum spanning 3830 to 4070~\AA~(which is mostly shown in
Fig. \ref{fig:spCD}) proved to be much more useful for abundance 
determination. The wavelength region of $\ld < 3870$~\AA~showed poor flux 
detection statistics and it was not used in the fit. The group of Fe, Mg and 
V lines at $\ld 3878$~\AA~were fitted reasonably well, although the blue edge 
has a slightly different slope. The Balmer line at $\ld 3889$~\AA~is somewhat 
deeper in the synthetic spectrum, but this may be caused by a not very 
detailed treatment of this broad line in the latter. The lines of Fe-peak 
elements between $\ld 3890$ and $\ld 3910$~\AA~can be fitted perfectly. 
However, although the depth of the Fe lines in the 3918--3920~\AA~region is 
well reproduced, there seems to be a problem with the continuum level, which 
shows a dip in the observed spectrum. 

The fit to the Ca\,{\sc ii}~K ($\ld 3933.7$~\AA) and H ($\ld 3968.5$~\AA) line 
profiles is very good and was useful for determining the rotational velocities 
along with some strong Fe lines. From the Ca\,{\sc ii}~H line to $\ld 4040$~\AA~the 
fit to the observed absorption lines is very satisfactory in such a way that 
the model and the observations are hardly distinguishable in most parts. It is 
only between $\ld 4020$ and $\ld 4028$~\AA~that a slight underabsorption 
is present in the observed spectrum, but it seems to be related to some 
continuum effect. In the remaining zone of the spectrum, we will only point 
out a couple of features besides the general good agreement. The Fe\,{\sc i} line 
profile at $\ld 4046$~\AA~presents a subestimated absorption, especially for 
the primary component. This behaviour, not observed in any other strong line, 
appears to be caused by the contribution of another absorption source. This 
is the strongest Fe\,{\sc i} line in this region, and some amount of interstellar 
absorption at $v_{\rm r}\approx0$ may be present so that it would be 
superimposed on the primary component line. The check of such a hypothesis 
would require the analysis of some spectra of the same region at different 
phases. Finally, the absorption Fe\,{\sc i} feature around $\ld 4055$~\AA~is 
overestimated in the synthetic spectrum while another quite strong Fe\,{\sc i} line 
at $\ld 4064$~\AA~is well reproduced.

In conclusion, the agreement between the observed and the synthetic spectrum
computed with the parameters listed in Table \ref{tab:CDTau} is good, as
shown in Fig. \ref{fig:spCD}. In some features, the slight flux underestimates 
may be caused by the poor $S/N$ of the observed spectra, by some deficiencies 
in the synthetic spectra or, more unlikely, in the model atmosphere fluxes. 

\section{The visual companion: CD~Tau~C} \label{cdc}

Radial velocity measurements of CD~Tau~C were also obtained with the 
CORAVEL scanner. In spite of the faintness of this star, near the limiting
magnitude of the instrument, a very clear correlation dip was present due to
the closeness of its spectral type to that of the mask (K2~III). The four
measurements lead to a mean $v_{\rm r}=-27.8\pm0.8$~\kms, which should 
be compared to the systemic velocity of the eclipsing system, $\gamma=
-29.3\pm0.3$~\kms, in order to assess the possibility of a physical link. As
it is seen, the agreement is fair. Moreover, the magnitude difference between 
CD~Tau~AB and CD~Tau~C is compatible with what is expected for a K-type star 
located at the same distance as the eclipsing pair. 

From Hipparcos measurements, the distance to CD~Tau is known to 
be 73$\pm$9~pc, which means that an angular separation of 9.98~arcsec can be 
translated into 730~AU. Assuming the total mass of the system to be around 
3.7~\Msol~(0.9~\Msol~is adopted for CD~Tau~C) and the current 
distance to be the semi-major orbital axis, Kepler's Third Law leads us to 
an orbital period of $\approx \!10\,000$~yr and a maximum expected radial
velocity difference of $\approx \!2.1$~\kms. Such a wide system seems not 
to be capable of surviving any weak encounter with another object. 
Thus, although it cannot be assured that CD~Tau~AB and CD~Tau~C are in orbit 
around each other, there exist evidences suggesting that they actually
form a common pair. 

The determination of the largest possible number of parameters for the visual 
component may be extremely useful when evaluating the model predictions. 
Unfortunately, neither the precision of the Str\"omgren photometric indices 
nor the intrinsic accuracy of the calibrations is enough for a reliable 
determination of the effective temperature of CD~Tau~C. Moreover, no IR 
observations could be collected due to the strong contamination caused by the 
bright companion. Thus, no reliable temperature estimation was possible from 
a photometric point of view.

However, an spectroscopic temperature determination of CD~Tau~C was performed 
via synthetic spectrum fitting. Along with the spectra of CD~Tau~AB, two 
spectra of its faint visual companion were also secured, with identical 
instrument configurations. The $S/N$ ratios of the spectra were of the order 
of 4 (the star is about 3~mag fainter), so that any possibility of abundance 
determination was ruled out. However, since there are strong evidences that 
CD~Tau~AB and CD~Tau~C are a physical system, we can reasonably assume that 
they share the same chemical composition. Thus, the spectra can provide a 
valuable estimation of the effective temperature of this early K-type star, 
for which there is no other clue. 

The spectrum fitting was performed in a similar way as for CD~Tau~AB, but 
taking into account the single nature of the star. The surface gravity was
adopted as 4.5~dex, the radial velocity as $-28$~\kms~(from CORAVEL 
observations) and the same chemical composition as CD~Tau~AB. Synthetic 
spectra were computed from Kurucz models at 250~K intervals between 4750~K 
and 5500~K and compared to the observed spectrum. In many zones, the choice 
of the best-fitting temperature was not clear due to the noise, but there were
two regions (around Ca~{\sc II}~H line, which shows some chromospheric emission,
and around $\ld 4045$~\AA, with a strong Fe~{\sc I} line) that allowed us to 
adopt 5250~K as the effective temperature that yields the best fit. The 
error of this determination is estimated to be around 200~K. 

Further information can be extracted from the magnitude difference between the 
eclipsing pair and the visual component. In the Hipparcos catalogue, the 
following values of the $Hp$ magnitude are listed for CD~Tau~AB and CD~Tau~C:
\[Hp_{\rm AB}=6.871\pm0.007 \hspace{1.5cm} Hp_{\rm C}=9.932\pm0.110\]
that lead to a magnitude difference $\Delta Hp=Hp_{\rm C}-Hp_{\rm AB}=
3.06\pm0.12$~mag. From $\Delta Hp$ we derive:
\[\Delta M_{\rm bol}=\Delta Hp+\Delta BC(Hp) \]
where $BC(Hp)$ is the bolometric correction for the $Hp$ passband. 
Straightforward transformations lead to: 
\[\Delta \log (\frac{L}{\mbox{\Lsol}})=-\frac{\Delta Hp+\Delta BC(Hp)}{2.5}\]
From the BC calibration of Cayrel et al. (1997) we adopted $BC(Hp)_{\rm AB}=
-0.15\pm0.05$ and $BC(Hp)_{\rm C}=-0.35\pm0.05$, leading to a $\Delta BC(Hp)=
-0.20\pm0.07$. These values and their estimated errors can be combined to 
derive $\Delta \log (L/\mbox{\Lsol})= -1.14\pm0.06$. 

The absolute luminosities of CD~Tau~A and B can be computed from their radii
and effective temperatures and we obtain:
\[\log (L/\mbox{\Lsol})_{\rm A}=0.63\pm0.02 \hspace{.5cm}
\log (L/\mbox{\Lsol})_{\rm B}=0.52\pm0.02\]
that can be combined to compute the total luminosity of the eclipsing pair, 
which comes out to be $\log (L_{\rm AB}/\mbox{\Lsol})=0.87\pm0.02$.
We finally obtain the luminosity of CD~Tau~C that is: 
\[\log (L_{\rm C}/\mbox{\Lsol})=-0.27\pm0.06\]
Also, from the effective temperature and the luminosity, the radius can be 
easily computed and we derive $R_{\rm C}=0.89\pm0.09$~\Rsol. All these values 
are also in agreement with the results obtained when adopting the magnitude 
difference listed in Wood (1976).

\begin{figure}
\leavevmode
\epsfxsize=\hsize
\epsfbox{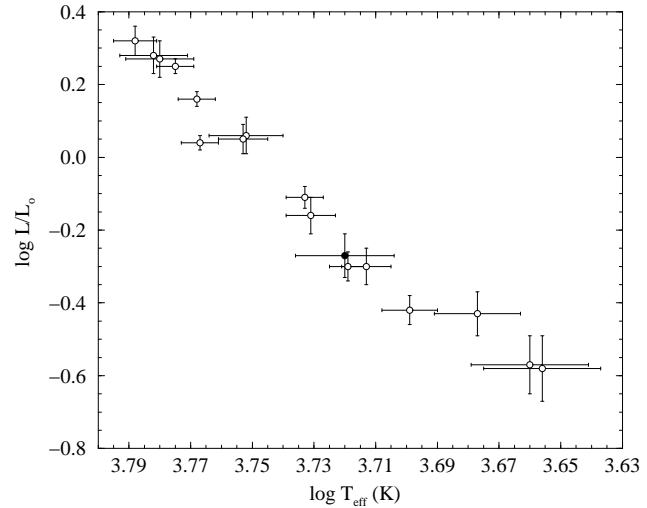}
\caption{H-R diagram of stars with masses comprised between 0.75 and
1.25~\Msol. Empty symbols are primary and secondary components of eclipsing
binaries, whereas the filled symbol represents CD~Tau~C.}
\label{fig:figHRLM}
\end{figure}

For illustrative purposes, we present in Fig. \ref{fig:figHRLM} an H-R 
diagram showing the position of CD~Tau~C and a sample of well-studied 
eclipsing binaries in the mass range $0.75<(M/\mbox{\Msol})<1.25$, taken
from the compilation of Clausen et al. (1999). As it can be seen, CD~Tau~C 
is located within the distribution described by all the eclipsing binary 
components.

\section{Comparison with evolutionary model predictions}

The comparison of model predictions and the observed properties of CD~Tau~AB
was done under the assumption that the evolutionary models should be able 
to fit an isochrone to both components of the system for a certain chemical 
composition. We made use of the evolutionary models of CG (Claret 1995, 1997; 
Claret \& Gim\'enez 1995, 1998) that constitute a grid with several $(Z,Y)$ 
values. The actual isochrone fitting was done by means of an interpolation
routine that uses the masses, radii and effective temperature of the
components and yields the best-fitting values for the age and the chemical 
composition $(Z,Y)$ of the system (among other parameters). Thus, both $Z$ 
(initial metallicity) and $Y$ (initial helium abundance) were treated as free 
parameters. The computational details of this algorithm will be provided in
a forthcoming paper (Ribas et al. 1999).

\begin{figure}
\leavevmode
\epsfxsize=\hsize
\epsfbox{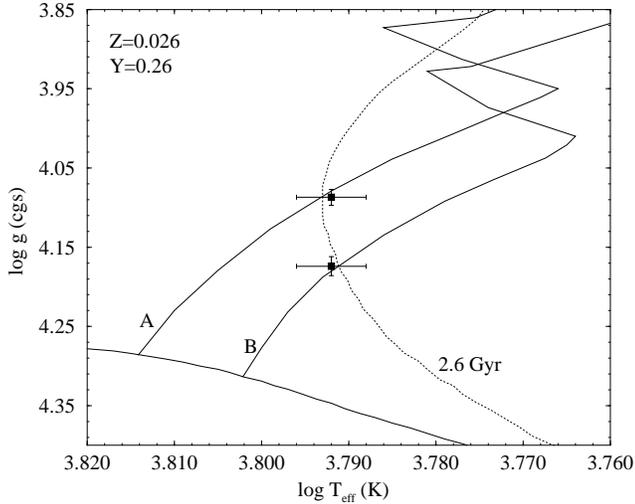}
\caption{$\g-\lte$ representation of evolutionary tracks and isochrone for 
CD~Tau~A and B. They are computed for the best-fitting value of the chemical 
composition by using the evolutionary models of CG (see text for references).}
\label{fig:evCD}
\end{figure}

The evolutionary models predict an age of $2.6\pm0.3$~Gyr and best-fitting
chemical composition values of $Z=0.026\pm0.007$ and $Y=0.26\pm0.04$. Indeed, 
the metal abundance derived from the isochrone fitting is in good agreement 
with the spectroscopic metallicity determination: $[Fe/H]= +0.08\pm0.15$
or $Z=0.023\pm0.008$. A $\g-\lte$ representation of CD~Tau~A and B, together
with the evolutionary tracks and isochrone computed for the best-fitting
chemical composition, is presented in Fig. \ref{fig:evCD}.

Considering the physical relation between CD~Tau~AB and CD~Tau~C, a common 
origin, both in time and composition, can be assumed. This implies that the 
same isochrone should simultaneously fit all three stars at an age of about 
2.6~Gyr. Due to the presumably large mass difference between the components, 
it is advisable to correct this estimated age by taking into account the 
different lifetime of the stars in the pre-main sequence (PMS) phase. The 
evolutionary tracks published by D'Antona \& Mazzitelli (1994) indicate that 
the difference in duration of the PMS phase (the mass of CD~Tau~C is
supposed to be 0.8-0.9~\Msol) is about 0.1~Gyr. So, the age of CD~Tau~C 
has to be corrected to a value of 2.5~Gyr when comparing to post-ZAMS 
evolutionary models.

\begin{figure}
\leavevmode
\epsfxsize=\hsize
\epsfbox{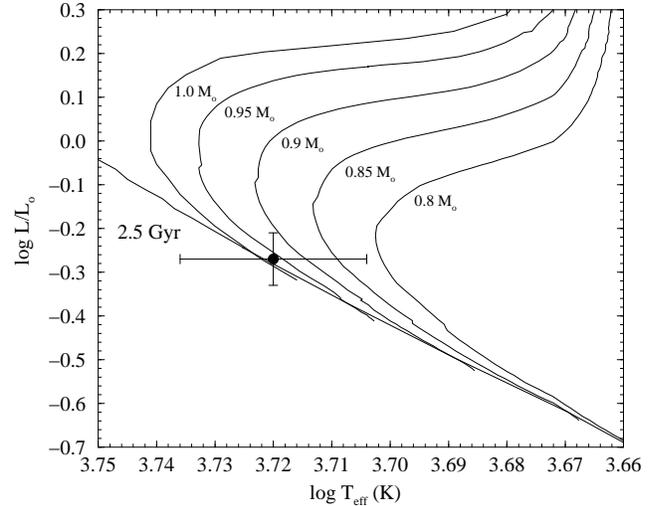}
\caption{H-R diagram showing the position of CD~Tau~C and evolutionary 
tracks for initial masses from 0.8 to 1.0~\Msol. A 2.5~Gyr isochrone (the 
age of CD~Tau~C) is also shown in the plot. The chemical composition was 
fixed at $Z=0.026$ and $Y=0.26$.}
\label{fig:figCDC}
\end{figure}

In order to compare the observed data with the model predictions, evolutionary 
tracks of initial masses 0.8, 0.85, 0.9, 0.95 and 1.0~\Msol~for the adopted 
chemical composition were kindly provided by Dr. A. Claret. They were computed 
with the same input physics of the CG evolutionary models. A $\lte-\log L$ plot 
of the evolutionary tracks and the observed values for CD~Tau~C is presented 
in Fig. \ref{fig:figCDC}. A 2.5~Gyr isochrone is also plotted. As it can be 
seen, the theoretical predictions and the observational data show an excellent 
agreement, well within the error bars. Notice that no free parameters remain, 
since the chemical composition of the star is also known. The best-fitting 
parameters that the models predict for an age of 2.5~Gyr are a mass of about 
$M=0.99$~\Msol~and a $\g$ of $4.54$~dex.

Further refinement of the effective temperature of CD~Tau C (by means of e.g. 
high-accuracy IR measurements of the visual component alone) may provide a 
more stringent test to the quality of the evolutionary models. What seems 
clear at this point, is that the systematics that Popper (1997) and Clausen 
et al. (1999) pointed out are not observed in CD~Tau~C. Indeed, the stellar 
parameters are very accurately described by the isochrone that fits the more 
massive components.

\section*{Acknowledgments}

E. Masana, F. Comer\'on, E. Oblak and S. Udry are thanked for their kind 
collaboration in performing and reducing the observations at Carlos S\'anchez 
Telescope, Mount Stromlo Observatory and Observatoire d'Haute Provence. A. 
Claret is acknowledged for making stellar evolutionary models for low-mass 
stars available to us. We also thank J.D. Pritchard for providing the latest 
version of the WD program used for light curve modelling, as well as several 
algorithms and routines. This work was supported by the Spanish CICYT under 
contract ESP97-1803. I.R. also acknowledges the grant of the {\em Beques
predoctorals per a la formaci\'o de personal investigador} by the CIRIT
(Generalitat de Catalunya)(ref. FI-PG/95-1111).

\bsp

\label{lastpage}


\begin{thebibliography}{99}
\bibitem{b1} Alonso A., Arribas S., Mart\'{\i}nez-Roger C., 1996, A\&AS, 
        117, 227
\bibitem{b2} Andersen J., 1991, A\&AR, 3, 91
\bibitem{b3} Baranne A., Mayor M., Poncet J.L., 1979, Vistas in Astronomy,
        23, 279
\bibitem{b4} Blackwell D.E., Petford A.D., Haddock D.J., Arribas S., Selby 
        M.J., 1990, A\&A, 232, 396
\bibitem{b5} Cayrel R., Castelli F., Katz D., Van't Veer C., G\'omez A., 
        Perrin M.-N., 1997, in Battrick B., ed., Hipparcos Venice 1997. 
        ESA SP-402, p. 433
\bibitem{b6} Cayrel de Strobel G., 1985, in Calibration of fundamental 
        stellar quantities. D. Reidel Publishing Co., Dordrecht, p. 137
\bibitem{b7} Chambliss C.R., 1992, PASP, 104, 663
\bibitem{b8} Clausen J.V., Helt B.E., Olsen E.H., 1999, in Gim\'enez A., 
        Guinan E.F., Montesinos B., eds., Theory and Tests of Convection 
        in Stellar Structure. ASP Conference Series, Vol. 173, in press
\bibitem{b9} Claret A., 1995, A\&AS, 109, 441
\bibitem{b10} Claret A., 1997, A\&AS, 125, 439
\bibitem{b11} Claret A., 1998, A\&AS, 131, 395
\bibitem{b12} Claret A., Gim\'enez A., 1995, A\&AS, 114, 549
\bibitem{b13} Claret A., Gim\'enez A., 1998, A\&AS, 133, 123
\bibitem{b14} Crawford D.L., 1975, AJ, 80, 955
\bibitem{b15} D'Antona F., Mazzitelli I., 1994, ApJS, 90, 467
\bibitem{b16} Duquennoy A., Mayor M., Halbwachs J.-L., 1991, A\&AS, 88, 281
\bibitem{b17} Edvardsson B., Andersen J., Gustafsson B., Lambert D.L., 
        Nissen P.E., Tomkin J., 1993, A\&A, 275, 101
\bibitem{b18} ESA 1997, The Hipparcos and Tycho Catalogues, ESA SP-1200
\bibitem{b19} G\"ulmen \"O., Ibano\v{g}lu C., G\"ud\"ur N., Bozkurt S., 
        1980, A\&AS, 40, 145
\bibitem{b20} Hauck B., Mermilliod M., 1998, A\&AS, 129, 431
\bibitem{b21} Jordi C., Galad\'{\i}-Enr\'{\i}quez D., Trullols E., 
        Lahulla F., 1995, A\&AS, 114, 489
\bibitem{b22} Kholopov P.N. et al., 1987, General Catalogue of Variable
        Stars, 4th. edition. Nauka, Moscow
\bibitem{b23} Kurucz R.L., 1991, in Crivellari L. et al., eds., Stellar 
        Atmospheres: Beyond Classical Models. Kluwer, Dordrecht, p. 441
\bibitem{b24} Kurucz R.L., 1994, CD-ROM No 19
\bibitem{b25} Lehmann-Filh\'es R., 1894, Astron. Nachr., 163, 17
\bibitem{b26} Masana E., 1999, PhD thesis, Univ. de Barcelona, in preparation
\bibitem{b27} Milone E.F., Stagg C.R., Kurucz R.L., 1992, ApJS, 79, 123
\bibitem{b28} Milone E.F., Stagg C.R., Kallrath J., Kurucz R.L., 1994, 
        BAAS, 184, 06.05
\bibitem{b29} Napiwotzki R., 1998, private communication
\bibitem{b30} Pols O.R., Schr\"oder K.-P., Hurley J.R., Tout C.A., 
        Eggleton P.P., 1998, MNRAS, 298, 525
\bibitem{b31} Popper D.M., 1971, ApJ, 166, 361
\bibitem{b32} Popper D.M., 1997, AJ, 114, 1195
\bibitem{b33} Ribas I. et al., 1999, MNRAS, in preparation 
\bibitem{b34} Russo G., Milano L., D'Orsi A., Marcozzi S., 1981, Ap\&SS, 
        79, 359
\bibitem{b35} Srivastava J.B., 1976, Ap\&SS, 40, 15
\bibitem{b36} Underhill A.B., 1966, The Early Type Stars, p. 127
\bibitem{b37} Wilson R.E., Devinney E.J., 1971, ApJ, 166, 605 (WD)
\bibitem{b38} Wood D.B., 1976, AJ, 81, 855
\end{thebibliography}
\end{document}